\documentclass[letterpaper]{article} % DO NOT CHANGE THIS
\usepackage{aaai25}  % DO NOT CHANGE THIS
\usepackage{times}  % DO NOT CHANGE THIS
\usepackage{helvet}  % DO NOT CHANGE THIS
\usepackage{courier}  % DO NOT CHANGE THIS
\usepackage[hyphens]{url}  % DO NOT CHANGE THIS
\usepackage{graphicx} % DO NOT CHANGE THIS
\def\UrlFont{\rm}  % DO NOT CHANGE THIS
\usepackage{natbib}  % DO NOT CHANGE THIS AND DO NOT ADD ANY OPTIONS TO IT
\usepackage{caption} % DO NOT CHANGE THIS AND DO NOT ADD ANY OPTIONS TO IT
\usepackage{algorithm}
\usepackage{algorithmic}
\usepackage{comment}
\usepackage{fontawesome5}
\usepackage{booktabs}
\usepackage[table]{xcolor}
\usepackage{tabularx}
\usepackage[most]{tcolorbox}

\usepackage{newfloat}
\usepackage{listings}
\DeclareCaptionStyle{ruled}{labelfont=normalfont,labelsep=colon,strut=off} % DO NOT CHANGE THIS
\floatstyle{ruled}
\newfloat{listing}{tb}{lst}{}
\floatname{listing}{Listing}
\title{Beyond Technocratic XAI: The Who, What \& How in Explanation Design}
\author{
    Ruchira Dhar\textsuperscript{\rm 1},
    Stephanie Brandl\textsuperscript{\rm 2},
    Ninell Oldenburg\textsuperscript{\rm 3}, 
    Anders Søgaard\textsuperscript{\rm 1,3}
    }
\affiliations{
    \textsuperscript{\rm 1}  Department of Computer Science, University of Copenhagen \\
    \textsuperscript{\rm 2} Center for Social Data Science, University of Copenhagen \\
    \textsuperscript{\rm 3} Department of Philosophy, University of Copenhagen \\
    rudh@di.ku.dk, stephanie.brandl@sodas.ku.dk, niol@hum.ku.dk, soegaard@di.ku.dk
}

\begin{document}

\maketitle

\begin{abstract}

The field of Explainable AI (XAI) offers a wide range of techniques for making complex models interpretable. Yet, in practice, generating meaningful explanations is a context-dependent task that requires intentional design choices to ensure accessibility and transparency. This paper reframes explanation as a situated design process—an approach particularly relevant for practitioners involved in building and deploying explainable systems. Drawing on prior research and principles from design thinking, we propose a three-part framework for explanation design in XAI: asking \textit{Who} needs the explanation, \textit{What} they need explained, and \textit{How} that explanation should be delivered. We also emphasize the need for ethical considerations, including risks of epistemic inequality, reinforcing social inequities, and obscuring accountability and governance. By treating explanation as a sociotechnical design process, this framework encourages a context-aware approach to XAI that supports effective communication and the development of ethically responsible explanations.
\end{abstract}

\section{Introduction}
The field of Explainable AI (XAI)  offers a wide array of methods to interpret and communicate the decisions made by complex models. Most works, however, focus on developing novel methods in the space \cite{ribeiro2016should, lundberg2017unified}, creating taxonomies to make users aware of different methods \cite{10.1145/3236009, arrieta2020explainable}, and creating scientific frameworks for comparing and evaluating methods \cite{doshi2017towards, vilone2021notions}. However, when deployed in practice, producing an explanation is a context-sensitive process that requires careful thought and design \cite{200027, kim2023designing, 10.1145/3613905.3651047, wolf2019explainability, wolf2020designing}. In domains like healthcare, finance, or criminal justice, explanations are not just tools for understanding but mechanisms for transparency, accountability, and trust-building. When misaligned with user needs or social context, explanations can mislead and reinforce existing inequities \cite{gerlings2021explainable, rane2023explainable, chaudhary2024explainable}. This leads us to an important question: \textit{how could someone best approach the task of generating an explanation in context?} While existing research highlights the need for human-centric XAI development \cite{doi:10.1126/scirobotics.aay4663, 10.1145/3491101.3503727, ehsan2024xai} and design awareness for bridging sociotechnical gaps \cite{ehsan2023charting, ehsan2024seamful}, there remains a gap in suggesting comprehensive end-to-end usable frameworks that can be easily applied to the design of an XAI process across deployment contexts.

What is needed is not just a toolbox of techniques, but a structured way to think about the explanation process itself, from intention to implementation. For XAI practitioners working to translate complex model behavior into accessible and trustworthy explanations, this shift is especially critical \cite{das2020opportunities, asghari2021explain, munn2022explainable}. To illustrate this, let us consider a scenario in which an XAI practitioner is tasked with explaining how a machine learning system detects early signs of a disease.  Before designing the explanation, they must ask:

\begin{itemize}
    \item \textbf{WHO} is going to use the explanation? A doctor or a patient? Their clinical background will determine what kind of explanation is appropriate.
    \item \textbf{WHAT} needs to be explained? Should the explanation give features that influenced the model’s decision? Or the false positive rate of all predictions in general? The depth and scope of the content depend on the user's role and decision-making needs.
    \item \textbf{HOW} should the explanation be delivered? Should it be accompanied by a brief textual summary or a confidence score? What kind of language should explanations use? The delivery format must align with the user’s workflow and cognitive preferences.
\end{itemize}

Moreover, once the explanation is delivered, an important question arises: does it support responsible and fair decision-making? Scholars have increasingly emphasized that explanation is not ethically neutral—it involves decisions about who gets to understand a system, what kind of knowledge is privileged, and how accountability is distributed \cite{mcdermid2021artificial}. In high-stakes domains like healthcare or finance, ethical adequacy means clarity on one hand but also enabling accountability and trust on the other \cite{gerlings2021explainable}. Such considerations can have a significant impact on how an explanation is designed for users within different contexts.

\begin{figure*}[ht!]
    \centering
    \includegraphics[width=0.8\linewidth]%{AnonymousSubmission/LaTeX/Fig1_v3.pdf}
    {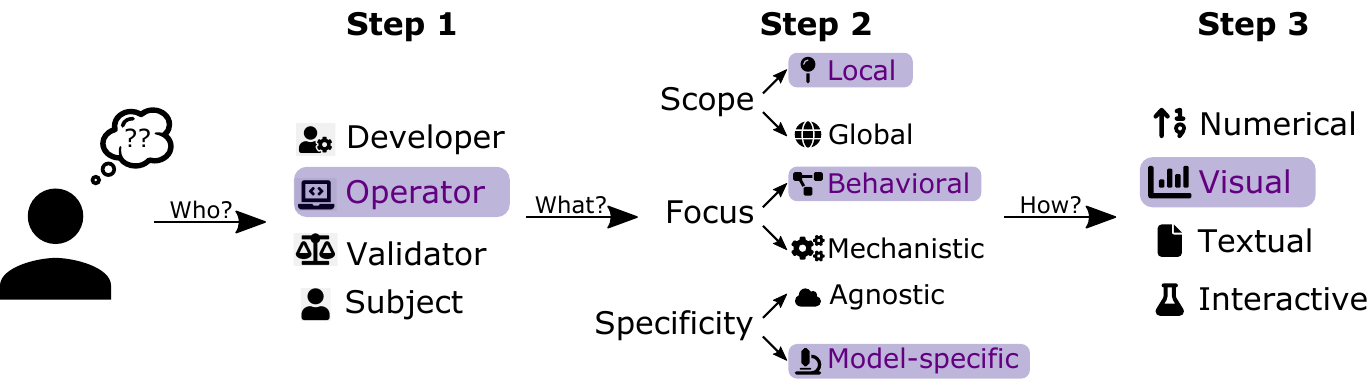}
    \caption{An Overview of our proposed XAI Design Process}
    \label{fig:overview}
\end{figure*}

This simplistic example captures a key point: explanations are always for someone, about something, and must be designed accordingly. Taking inspiration from design thinking \cite{brown2008design, brown2015everyone} and research advocating human-centric explanations \cite{wang2019designing, ehsan2020human, ehsan2021expanding,frank2024paradigm,oldenburg2025navigating}, we propose a framework for treating explanation as a sociotechnical practice. This also complements recent work in XAI evaluation that emphasizes the importance of user-centered metrics and human factors \cite{10.1145/3313831.3376590, ehsan2024xai, frank2024paradigm}. While our methodological discussion primarily focuses on widely used XAI approaches in natural language processing (NLP), the framework we present is adaptable to a broad range of AI systems and application contexts.

In  \textit{Background \& Related Work}, we discuss some essential terminology used in the paper. We also discuss related work in the field that has served as inspiration and is relevant to the premise of our work. In \textit{Design Process in XAI}, we address each step of the process individually and provide key consideration points in each step. In \textit{Ethics in XAI},  we highlight some key considerations for reflecting on ethical aspects of the decisions made in the design process.  In \textit{Reframing XAI as Design Process}, we summarize the process so far, and in \textit{Design Process in Action}, we also take on a concrete example for illustrating the process with an imagined use-case. We give a step-wise instantiation of our design process, and in each step, provide ethical reflections on design choices made during the process. 

By shifting from method-centric thinking to process-oriented design, we aim to support practitioners in building explanation systems that are not only technically robust but also contextually appropriate and ethically sound.

%%%%%%%%%%%%%%%%%%%%%%%%%%%%%%%%%%%%%%%%%%%%%%%%%%%%%%%%%%%%%%%%%%%%%%%%%%%%%%%%%%%%%%%%%%%%%%%%%%%%%%%%%%%%%%%%%%%%%%%%%%%%%%%%%%%%%%%%%%

\definecolor{lightgray}{gray}{0.95}

\begin{table*}[t]
\centering
\rowcolors{2}{lightgray}{white}
\begin{tabular}{@{}m{2.8cm} m{3cm} m{1.7cm} m{1.7cm} m{6.5cm}@{}}
\toprule
\textbf{Stakeholder Type} & \textbf{Demographic} & \textbf{Domain Knowledge} & \textbf{Technical Knowledge} & \textbf{Goal} \\
\midrule
\faLaptopCode\ \textbf{Developers} & Researchers \newline Engineers  & Low & High & Analyze, debug, and improve model behavior \\
\faUserCog\ \textbf{Operators} & Doctors \newline Bankers & High & Low & Use model outputs to support decisions \\
\faBalanceScale\ \textbf{Validators} & Auditors \newline Compliance Officers & Medium & Medium & Assess fairness, robustness, and regulatory compliance \\
\faUser\ \textbf{Subjects} & Patients \newline Customers & Low & Low & Understand, contest, or build trust in decisions that affect them \\
\bottomrule
\end{tabular}
\caption{Answering the \textit{WHO}: Categorizing stakeholders to support explanation design in XAI}
\label{tab:stakeholders}
\end{table*}

\section{Background \& Related Work}
\label{sec:background}

\paragraph{Terminology}

Explainability is commonly understood as the ability to articulate why a model produced a given output in a way that is accessible to human users, often emphasizing post hoc techniques that support understanding of model outputs \cite{arrieta2020explainable, wiegreffe2022reframing, dwivedi2023explainable, kunz2024properties}. Interpretability, by contrast, is sometimes defined more narrowly as the degree to which a model’s internal mechanics can be directly understood, typically by experts and often through inherently transparent model architectures or diagnostic tools \cite{doshi2017towards}. More recently, Mechanistic Interpretability (MI) has emerged as an umbrella term for another category of interpretability methods that are mostly aimed at interpreting model outputs in terms of their internal representations and architectures \cite{bereska2024mechanistic}. The  ``traditional” XAI community and the more recent MI community so far operate mostly in parallel. \citet{saphra-wiegreffe-2024-mechanistic} made a first step towards disentangling the two communities.  While the distinction of explainability and interpretability has guided parts of the literature \cite{miller2019explanation}, both terms are often used interchangeably in the literature. 

In this paper, we do not maintain a strict separation between explainability and interpretability. Instead, we treat explanation as a concept that spans the entire pipeline from interrogating black-box models to communicating justifications that are meaningful to stakeholders. This includes both technical introspection and communicative delivery.

\paragraph{XAI Methods and Their Limits}
A wide range of technical methods for XAI have been proposed, including feature attribution (e.g., Integrated Gradients, Layer-wise Relevance Propagation), surrogate models (e.g., LIME, SHAP), and example-based explanations (e.g., counterfactuals). These methods aim to increase transparency but often fall short in practice due to their reliance on technical assumptions that do not align with users' mental models or real-world constraints \cite{alvarez2018robustness}. Moreover, they are designed without considering the interpretive needs or capacities of \textit{specific} users. Recent work in mechanistic interpretability \cite{saphra-wiegreffe-2024-mechanistic} offers a complementary view focused on opening the ``black box" by analyzing internal model structure, but largely remains developer-oriented. In this paper, we will try to close this gap by including the most prominent explainability approaches into our design framework.

\paragraph{Human-Centered \& Design-Oriented XAI}
Recognizing the limits of purely technical explanations, researchers have explored user-centered and participatory approaches to XAI \cite{abdul2018trends, kaur2020interpreting, 10.1145/3491101.3503727, ehsan2020human, ehsan2023charting, ehsan2024xai, ehsan2024seamful}. These include user studies on explanation efficacy \cite{jacovi2023diagnosing}, comparing model explanations to either human attention in the form of eye-tracking \cite{brandl2024evaluating} or human annotations \cite{jakobsen2023being}, and interactive explanation interfaces \cite{8807255, hohman2019gamut}. While there is some work on design-oriented XAI \cite{wolf2019explainability, wolf2020designing}, there is a lack of general systematic frameworks to guide explanation design across contexts. Our work builds on this tradition, proposing a principled design framework grounded in user-centric thinking.

\paragraph{Ethical Perspectives on Explanation}
Explanation is not a value-neutral activity but involves decisions about whose understanding matters, what kinds of knowledge are privileged, and how accountability is distributed \cite{binns2018fairness, selbst2019fairness}. Scholars in Science Technology Studies and Human-Computer Interaction (HCI) have highlighted the risks of epistemic injustice \cite{langton2010miranda}, design bias \cite{costanza2020design}, and abstraction traps \cite{selbst2019fairness} in XAI systems. Our work positions explanation as a sociotechnical process that must grapple with these ethical and normative tensions, offering a framework for designing explanations that are not only useful but ethical.

\begin{comment}
    
\definecolor{lightgray}{gray}{0.95}

\begin{table*}[t]
\centering
\caption{Stakeholder-Centric Classification of XAI Needs}
\label{tab:stakeholders}
\rowcolors{2}{lightgray}{white}
\begin{tabularx}{\textwidth}{@{}l X X X@{}}
\toprule
\textbf{Stakeholder Type} & \textbf{Demographic} & \textbf{Knowledge Level} & \textbf{Goal} \\
\midrule
\faLaptopCode\ \textbf{Developers} & ML researchers, data scientists, model engineers & Low Domain Knowledge \newline High Technical Knowledge & Understand model behavior, debug, and improve performance \\
\faUserCog\ \textbf{Operators} & Doctors, loan officers, analysts, moderators, customer service agents & High Domain Knowledge \newline Low Technical Knowledge & Make informed, context-specific decisions using model outputs \\
\faBalanceScale\ \textbf{Validators} & Auditors, compliance officers, internal reviewers, domain experts & Medium Domain Knowledge \newline Low Technical Knowledge & Assess fairness, robustness, accountability, and regulatory compliance \\
\faUser\ \textbf{Subjects} & Consumers, patients, applicants, citizens & Low Domain Knowledge \newline Low Technical Knowledge & Understand, contest, or build trust in decisions that affect them \\
\bottomrule
\end{tabularx}
\label{tab:stakeholders}
\end{table*}
\end{comment}

%%%%%%%%%%%%%%%%%%%%%%%%%%%%%%%%%%%%%%%%%%%%%%%%%%%%%%%%%%%%%%%%%%%%%%%%%%%%%%%%%%%%%%%%%%%%%%%%%%%%%%%%%%%%%%%%%%%%%%%%%%%%%%%%%%%%%%%%%%

\section{Design Process in XAI}

In this section, we introduce our main contribution: a human-centered framework for designing explanations in XAI, organized around three key questions: \textit{Who, What, and How} as outlined in Figure \ref{fig:overview}. This framework encourages a shift from focusing solely on algorithmic methods to considering the broader context in which explanations are created and used. It is also the basis for delineating the ethical accounts of explanation design in the next section.

\subsection{Answering the WHO}\label{sec:who}

The first question in explanation design is simple: \textit{Who is the explanation for?} Unlike most technical work in XAI that assumes a generic user, human-centered design demands that we attempt to identify and differentiate among the stakeholders engaged with or affected by an AI system \cite{miller2019explanation,decker2023thousand}.

Existing work has tried to come up with criteria for identifying stakeholders in the explainability space--ranging from chronological order in a traditional machine learning pipeline \cite{dwivedi2023explainable}, the need for explanations \cite{calderon-reichart-2025-behalf}, and legal considerations \cite{mansi2025legally}. For applied contexts, however, there are few requirements to be kept in mind: 

\begin{itemize}
    \item Reliance on professional titles can create false dichotomies where multiple stakeholder groups might have similar goals (e.g., developers, theorists, and data scientists can all have conflated goals of understanding the model better in general). 
    \item Stakeholder classification needs to be based on goals but also be fine-grained enough to identify and differentiate individual in real world scenarios. \citet{calderon-reichart-2025-behalf} clubs together significantly different stakeholders like enterprises and end-users (who can have different interests in XAI methods) while creating a theoretically vague class of ``society'' which potentially includes all stakeholders at large. 
\end{itemize}

We take inspiration from \citet{habibullah2024explainable} which proposes a user taxonomy based on a rigorous literature review of the field. In addition, we provide factors of consideration such as demographics, domain knowledge, technical knowledge, and goals that can help distinguish actors in real-world scenarios. Table~\ref{tab:stakeholders} summarizes four groups: Developers, Validators, Operators, and Subjects. This framing draws on both life cycle roles and interpretive purpose, while highlighting differences in access and accountability.

\begin{itemize}
    \item \textbf{Developers} require explanations to understand, debug, and refine model internals. Explanations must prioritize technical faithfulness and granularity.

    \item \textbf{Operators} use model outputs to support real-time decision-making in high-stakes settings. Explanations must be context-sensitive, actionable, and efficient.
    
    \item \textbf{Validators} such as auditors or compliance officers seek transparency to assess fairness, risk, or adherence to norms. This group can also include advocacy groups who have similar goals of ensuring fairness. Explanations here must be comprehensible, structured, and aligned with regulatory standards.
    
    \item \textbf{Subjects} are individuals impacted by model decisions. They may not interact directly with the model, but their rights and outcomes are shaped by it. Explanations here must be accessible to empower recourse.
\end{itemize}

It is important to note that the roles are not always mutually exclusive, but that a single individual may occupy multiple positions across the AI lifecycle depending on real-world contexts of deployment. However, our framework is meant to provide guidance in such situations instead of enforcing definitive and rigid dichotomies. By articulating the knowledge constraints and distinct goals of different stakeholders, this classification supports the design of explanation methods that are meaningfully aligned with their intended users.  For instance, a developer may take on the role of validator during model audits. Despite these overlaps, each role reflects a distinct mindset shaped by specific constraints, expectations, and forms of accountability. Designing effective explanations requires attending to these mindsets as meaningful user perspectives, not just organizational labels \cite{dhanorkar2021needs, ehsan2024xai}.

%%%%%%%%%%%%%%%%%%%%%%%%%%%%%%%%%%%%%%%%%%%%%%%%%%%%%%%%%%%%%%%%%%%%%%%%%%%%%%%%%%%%%%%%%%%%%%%%%%%%%%%%%%%%%%%%%%%%%%%%%%%%%%%%%%%%%%%%%%

\subsection{Answering the WHAT}

Once the relevant stakeholder has been identified, the next critical design decision is: \textit{What should be explained}? This question shapes not only the type of explanation to be provided but also its relevance, utility, and impact. In practice, it is not enough to generate technically sound explanations. Effective XAI must align with stakeholder goals, cognitive capacity, and operational context. An explanation that targets the wrong aspect of model behavior risks misleading its audience or becoming practically unusable.

To support better-aligned design decisions, we classify existing XAI methods along four axes that reflect their real-world implications \cite{madsen2022post,dwivedi2023explainable, calderon-reichart-2025-behalf}: (1) \textbf{explanation scope}, (2) \textbf{explanation focus}, (3) \textbf{model specificity}, and (4) \textbf{operational cost}. We see these dimensions as design-relevant levers that help match methods to the needs of different stakeholders, introduced earlier. Our discussion centers primarily on post hoc explanation methods typically applied to textual AI systems like LLMs \cite{wang2024history}, i.e., models that are typically black-box in nature and do not offer intrinsic or self-explaining capabilities \cite{loyola2019black}. Furthermore, these four design axes are broadly applicable and can be extended to guide explanation strategies across a wider range of AI models and domains, even though our emphasis lies on methods relevant to contemporary NLP systems.

\begin{table*}[t]
\label{tab:method-taxonomy}
\centering
\rowcolors{2}{lightgray}{white}
\begin{tabularx}{\textwidth}{@{}m{3.5cm} m{2.5cm} m{2.5cm} m{2cm} m{5.4cm}@{}}
\toprule
\textbf{Method} & \textbf{Scope} & \textbf{Focus} & \textbf{Specificity} & \textbf{Best-fit Stakeholders} \\
\midrule
Grad.-based Feat. Attrib.& \faMapPin\ Local & \faProjectDiagram\ Behavioral & \faMicroscope\ Specific & \faLaptopCode\ Developers, \faBalanceScale\ Validators\\
% SHAP & \faMapPin\ Local & \faProjectDiagram\ Behavioral & \faCloud\ Agnostic & \faArrowUp\ High & \faUserCog\ Operators, \faUser\ Subjects \\
% LIME & \faMapPin\ Local & \faProjectDiagram\ Behavioral & \faCloud\ Agnostic & \faEquals\ Medium & \faUserCog\ Operators \\
Surrogate Methods & \faMapPin Local/ \faGlobe Global & \faProjectDiagram\ Behavioral & \faCloud\ Agnostic & \faLaptopCode\ Developers, \faBalanceScale\ Validators \\
Counterfactuals & \faMapPin\ Local & \faProjectDiagram\ Behavioral & \faCloud\ Agnostic & \faUserCog\ Operators, \faBalanceScale\ Validators, \faUser\ Subj. \\
Probing & \faGlobe\ Global & \faProjectDiagram\ Behavioral & \faMicroscope\ Specific & \faBalanceScale\ Validators\\
% PDP & \faGlobe\ Global & \faProjectDiagram\ Behavioral & \faCloud\ Agnostic & \faArrowDown\ Low & \faLaptopCode\ Developers, \faBalanceScale\ Validators \\
Circuit Tracing & \faGlobe Global & \faCogs\ Mechanistic & \faMicroscope\ Specific  & \faLaptopCode\ Developers \\
Sparse Autoencoders & \faGlobe\ Global & \faCogs\ Mechanistic & \faMicroscope\ Specific & \faLaptopCode\ Developers \\
\bottomrule
\end{tabularx}
\caption{Answering the \textit{WHAT}: Taxonomy of explanation methods across design-relevant axes. We group methods here into categories. Gradient-based feature attribution includes methods like LRP, Integrated Gradients, Saliency Mapping, and Gradient $\times$ Input. SHAP and LIME fall under the umbrella of (local) surrogate models.}
\label{tab:methods}
\end{table*}

\paragraph{Explanation Scope}  
This axis refers to the level of generality at which a method operates. \textbf{Local methods} provide explanations for individual predictions. These are crucial in high-stakes applications where specific decisions must be justified, such as loan approvals or medical diagnoses. Popular local techniques include gradient-based methods like integrated gradients \cite{sundararajan2017axiomatic}, some surrogate methods like LIME \cite{ribeiro2016should}, or counterfactual methods \cite{dai2022counterfactual}.

In contrast, \textbf{Global methods} aim to explain the overall behavior of a model across many inputs. They are well-suited for understanding patterns the model has learned and are often useful in development or validation settings. Examples include some surrogate methods like SHAP \cite{lundberg2017unified} and global surrogate modelling \cite{apley2020visualizing, zhu2022fuzzy}, probing \cite{8807255}, circuit tracing \cite{olah2020zoom, conmy2023towards}, and sparse autoencoders \cite{sharkey2022taking}. Some other popular methods also include partial dependence plots \cite{friedman2001greedy, greenwell2017pdp}, individual conditional expectation \cite{goldstein2015peeking}, etc.

From a stakeholder lens, global methods are typically valuable for \textit{developers} and \textit{validators} who are focused on the model's general behavior, while local methods better serve \textit{operators} and \textit{subjects} who are concerned with case-specific outcomes.

\paragraph{Explanation Focus}  
Different explanation methods shed light on different aspects of a model. Are we illuminating which parts of the input were most influential on the model’s output, or are we uncovering the internal mechanisms, such as features, representations, or circuits, that the model uses to compute its decision? \textbf{Behavioral methods}, often referred to as ``post-hoc'' methods,  aim to explain how inputs influence outputs, often by observing model behavior under perturbations or through feature attribution \cite{retzlaff2024post, cesarini2024explainable}. These include gradient-based feature attribution methods like  Integrated Gradients \cite{sundararajan2017axiomatic}, but also Surrogate Models like SHAP \cite{lundberg2017unified} and LIME \cite{ribeiro2016should}, counterfactuals, and probing techniques.

By contrast, \textbf{mechanistic methods} attempt to uncover the internal structures, representations, or circuits that drive model behavior. This emerging paradigm, often associated with mechanistic interpretability \cite{olah2020zoom, nanda2023progress}, includes tools such as neuron-level attribution \cite{geiger2021causal}, sparse autoencoders \cite{elhage2022toy}, and circuit tracing techniques \cite{conmy2023towards}.

The choice of explanation focus should be guided by what the stakeholder needs to understand to achieve their goal. \textit{Developers}, for example, often benefit from mechanistic explanations that reveal how internal components such as neurons, layers, or circuits contribute to model behavior, enabling them to debug or refine the system. \textit{Validators}, such as auditors or fairness reviewers, may require both behavioral and mechanistic insights to assess robustness, compliance, or potential bias. For \textit{operators} and \textit{subjects}, however, behavioral explanations tend to be more intuitive and actionable, particularly example-based or counterfactual methods. They offer clear links between inputs, outputs, and possible alternatives. Crucially, this design choice does not just affect what the explanation reveals, but it has important consequences for how the explanation is interpreted and used in sensitive decision-making scenarios \cite{jesus2021can, vale2022explainable}.

\paragraph{Model Specificity}  
Explanation methods also differ significantly in how much internal access they require to the model being explained and how specific any explanations are to the model at hand. This axis, often referred to as model specificity \cite{darias2021systematic, sandu2022comparing, letrache2023explainable}, is a key aspect in determining the relevance of explanations. 

The \textbf{model-specific} approaches rely on internals like gradients \cite{sundararajan2017axiomatic}, representations for probing \cite{8807255}, attention weights for circuits \cite{conmy2023towards}, or neuron activations and features in sparse autoencoders \cite{massidda2023causal}. These methods are tightly coupled to the model’s architecture and learning dynamics, allowing for high-resolution insights into how a particular model processes information. For example, techniques like Integrated Gradients or LRP compute feature attributions based on backpropagated gradients, while transformer-specific methods can trace attention flow or patch activations to localize functionality. Since model specific approaches can access the internal model internal components, the explanations are grounded in the model’s internal structure and behavior \cite{calderon-reichart-2025-behalf}. However, they require full access to the model and its internals, which is not always feasible when the model is closed-source or where privacy concerns are paramount \cite{spartalis2023balancing, ezzeddine2024privacy}.

On the other hand, \textbf{model-agnostic} methods treat the model as a black box, relying only on inputs and outputs. These include methods relying on surrogate models like LIME and (kernel) SHAP or counterfactuals. While more widely applicable and well-suited for settings where the underlying model is proprietary, model-agnostic methods may offer less faithful approximations and struggle with complex or highly non-linear models \cite{molnar2020general}.

From a design perspective, \textit{developers} and \textit{validators} working in trusted environments may prefer model-specific methods, while \textit{operators} and \textit{subjects}, who often face opaque third-party models, depend on model-agnostic approaches. For them, model-agnostic methods offer essential transparency, even if less granular. Selecting an explanation method, therefore, must align not only with what the model can support, but also with who the user is and what kind of interaction is realistically possible.

\paragraph{Operational Cost}  
Finally, practical deployment requires attention to the cost of explanation, both in terms of \textbf{time latency} and \textbf{compute load}. Perturbation-based methods like LIME \cite{ribeiro2016should} and kernel SHAP \cite{lundberg2017unified, roshan2022using} often involve hundreds of model queries, resulting in noticeable delays and high memory use. Gradient-based methods are lighter because they reuse model internals, but can become costly when explaining many inputs in batch or when dealing with massive models. Mechanistic methods are often the most computationally intensive, involving large-scale activation patching, neuron tracing, or circuit discovery that may require dedicated infrastructure \cite{luccioni2025bridging}.

These costs are not just a matter of engineering but affect who can realistically benefit from XAI. In low-latency environments, e.g., mobile traffic classification \cite{nascita2021xai} or anomaly detection \cite{gummadi2024xai}, lightweight behavioral methods may be the only viable option. In resource-constrained or sustainability-focused settings, compute-heavy techniques may be ethically or environmentally untenable \cite{stojkovic2024towards,luccioni2024power,luccioni2025bridging}. Thus, cost-conscious design is essential, and some recent work has made progress towards attempts at including cost considerations of XAI in model design \cite{jean2023cost}.

Cost considerations are a critical yet often overlooked aspect of explanation design in XAI. In general, \textit{developers} and \textit{validators} may accept higher cost for deeper analysis, but \textit{operators} and \textit{subjects} typically need lightweight, fast, and low-resource explanations.

\medskip

Taken together, these four axes provide a practical lens for selecting explanation methods that align with the diverse needs, capacities, and constraints of different stakeholders. Table~\ref{tab:methods} offers an illustrative mapping of commonly used XAI methods across three of the axes: scope, focus, and model specificity. We exclude operational cost from this table, as it is highly variable and depends on factors such as model architecture, input size, and the number of instances to be explained. Similarly, the stakeholder-method mappings should not be read as definitive recommendations; rather, they serve as indicative starting points.  The aim of this illustration is not to give prescriptive matches, but starting points for informed design decisions.

%%%%%%%%%%%%%%%%%%%%%%%%%%%%%%%%%%%%%%%%%%%%%%%%%%%%%%%%%%%%%%%%%%%%%%%%%%%%%%%%%%%%%%%%%%%%%%%%%%%%%%%%%%%%%%%%%%%%%%%%%%%%%%%%%%%%%%%%%%

\subsection{Answering the HOW}

\begin{table*}[t]
\label{tab:explanation-modalities}
\centering
\rowcolors{2}{lightgray}{white}
\begin{tabularx}{\textwidth}{@{}m{2cm} m{5cm} m{5.5cm} m{3.8cm}}
\toprule
    \textbf{Modality} & \textbf{Description} & \textbf{Example Tools / Techniques} & \textbf{Best-fit Stakeholders} \\
    \midrule

    \faSortNumericUp\ Numerical & Confidence scores, feature importances, statistical summaries & SHAP values, LIME values & \faLaptopCode\ Developers \\

    \faChartBar\ Visual & Plots, saliency maps, heatmaps, dashboards &  Grad-CAM Plots,  LIME visualizations & \faLaptopCode\ Dev., \faBalanceScale\ Val., \faUserCog\ Op. \\

    \faFile \ Textual & Explanatory text, simplified justifications, summaries & Natural language rationales, Rules, Counterfactuals, Chain-of-Thought Explanations & \faBalanceScale\ Val., \faUserCog\ Op., \faUser\ Subj. \\

    \faFlask\ Interactive & Exploratory tools allowing user interaction with model behavior & What-if tools, Causal simulators & \faUserCog\ Op., \faUser\ Subj. \\
\bottomrule
\end{tabularx}
\caption{Answering the \textit{HOW}: Typical Modalities of Explanation Delivery}
\label{tab:typesexp}
\end{table*}

The final design question becomes: \textit{How should the explanation be delivered}? But this question is more than a technical decision about format. It is an act of epistemic mediation, i.e., deciding how knowledge is shaped, prioritized, and framed for specific ends. As \citet{miller2019explanation} notes, explanations are not mere outputs but social performances: constructing narratives, eliciting trust or doubt, and reflecting assumptions about what kinds of understanding ``count.”

In this view, explanation delivery is fundamentally a design problem. It involves selecting the right \textbf{format}, \textbf{modality}, and \textbf{structure} to support cognitive alignment between the model and its users. Recent work also indicates that explanation format significantly influences user trust, comprehension, and reliance \cite{kulesza2013too, 10.1145/3313831.3376590, ehsan2024xai}. An explanation that is accurate but poorly delivered may be ignored, misunderstood, or even mistrusted. Effective explanation thus requires careful attention to how insights are surfaced.

Common explanation formats fall into four broad categories: \textbf{Numerical}, \textbf{Visual}, \textbf{Textual}, and \textbf{Interactive}. Each offers affordances as well as limitations, omissions, and implicit ideologies about the nature of understanding.

\paragraph{Numerical Explanations}  
These explanations use quantitative indicators such as scores, weights, or ranks to convey feature importance or model confidence. Common methods include gradient-based feature attribution like Integrated Gradients or surrogate methods like LIME, SHAP, etc. Numerical formats are compact, expressive, and precise, especially useful for \textit{developers} or \textit{validators} who are comfortable with abstract metrics. However, they assume statistical fluency and often lack narrative support, which makes them rather inaccessible to non-experts. Moreover, they often mask uncertainty under a veneer of objectivity, privileging technical legibility over narrative intelligibility. In contexts like healthcare or public services, this can reinforce epistemic asymmetries between technical experts and impacted individuals \cite{wolf2020designing}.

\paragraph{Visual Explanations} 
Visual methods leverage graphical encodings like heatmaps, saliency maps, and attention plots to illustrate what the model “sees” or attends to. Examples include Grad-CAM \cite{selvaraju2020grad}, or attention visualization in transformers \cite{belrose2023eliciting}. These methods offer immediate, intuitive understanding and are especially powerful for \textit{operators} and \textit{subjects} who benefit from fast glanceable insight. 

Note that visual explanations are often generated from the numerical explanations that are generated first. Thus, they can oversimplify or obscure causal mechanisms, and may lead users to overinterpret the output \cite{jain-wallace-2019-attention, lyu2024towards}. 

\paragraph{Textual Explanations} 
Textual explanations transform model reasoning into natural language, either as post-hoc rationalizations or integrated reasoning chains. Variants include \textit{explain-then-predict} \cite{lei-etal-2016-rationalizing,camburu2018snli}, \textit{predict-then-explain} \cite{marasovic-etal-2022-shot}, and \textit{chain-of-thought} prompting in LLMs. These formats excel at accessibility and narrative coherence, particularly for \textit{subjects} and non-technical stakeholders. However, natural language explanations may sacrifice faithfulness or omit uncertainty, creating a false sense of understanding if not critically designed \cite{agarwal2024faithfulness, doi:10.1126/scirobotics.aay4663}. 

As recent work shows, textual explanations are prone to hallucination, simplification, and selective emphasis \cite{agarwal2024faithfulness}. In general, language invites narrative closure and a premature sense of “knowing” that may conceal model fallibility or normative bias. When these explanations are deployed to justify high-stakes decisions (e.g., denial of loans), their rhetorical polish can dangerously undermine accountability rather than support it.

\paragraph{Interactive Explanations}  
Interactive formats empower users to engage with the explanation, exploring what-if scenarios, adjusting inputs, or navigating between explanation layers. Tools like Google’s \textit{What-If Tool} \cite{8807255} or \textit{Gamut} \cite{hohman2019gamut} blend visual, numerical, and textual elements into a responsive interface. Interactive systems offer flexible, layered access to explanation, letting users manipulate inputs, simulate counterfactuals, or explore uncertainty \citep{oldenburg2025navigating}. These are especially attractive to \textit{developers}, \textit{validators}, and skilled \textit{operators} navigating complex models.

\medskip

It’s important to note that these formats are not mutually exclusive and can also be combined. A single explanation can be translated across modalities, e.g., SHAP values rendered as bar charts, tables, or verbal summaries. The challenge hereby lies in choosing the right format for the user’s context, skill level, and goals.

In human-centered XAI, explanation delivery is consequential \cite{ehsan2024xai}. When poorly rendered, the same logic may confuse or mislead; when well-crafted, it can inform, empower, and build trust \cite{kulesza2013too}. \textit{Developers} often prefer low-level numerical or interactive formats that support debugging while \textit{validators} benefit from mixed-modal explanations that document decisions while revealing latent patterns \cite{naiseh2023different}. \textit{Operators} such as clinicians or analysts rely on fast, interpretable visual or interactive formats to make quick decisions \cite{scharowski2023exploring, brankovic2025clinician}. \textit{Subjects} require natural language or counterfactuals that connect model decisions to personal impact, often with minimal cognitive load \cite{ehsan2024xai}.

Finally, even within stakeholder groups, users vary. A senior developer may grasp low-level metrics, while a new team member may need higher-level structure. Moreover, users often suffer from the illusion of understanding, i.e., “good-enough” comprehension: believing they understand an explanation when their mental model is incomplete \cite{kaur2024interpretability, jacovi2023diagnosing}. To prevent this problem, it is important to consider how explanation formats not only shape what is explained but also what is understood.

%%%%%%%%%%%%%%%%%%%%%%%%%%%%%%%%%%%%%%%%%%%%%%%%%%%%%%%%%%%%%%%%%%%%%%%%%%%%%%%%%%%%%%%%%%%%%%%%%%%%%%%%%%%%%%%%%%%%%%%%%%%%%%%%%%%%%%%%%%

\section{Ethics in XAI}

Explanations in AI are often framed as a technical remedy for opacity, or as an add-on for transparency or trust. However, this framing obscures the ethical stakes. Decisions about the who, what, and how of explanations are not neutral, but shape users’ ability to know, challenge, and contest automated decisions \cite{stone2025legitimate}. In this section, we highlight three dimensions that demand ethical attention in the design of explainable AI systems: epistemic inequality, social inequality, and accountability.

\subsection{Epistemic Inequality}

Explanations are not distributed equally across stakeholders. Developers and validators typically gain access to rich model internals and tooling, while affected individuals such as patients, applicants, or defendants often receive vague rationales or simplified narratives \cite{crawford2016ai, ananny2018seeing}. This selective legibility reinforces what Fricker calls \textit{epistemic injustice} \cite{fricker2017evolving}, in which marginalized groups are denied full participation in knowledge practices that impact them. Recent work in AI ethics and design has highlighted how information asymmetry in ML systems can deepen structural exclusion \cite{edwards2017slave, veale2017fairer, veale2019administration}. Explanation frameworks must grapple with this asymmetry, not only by identifying stakeholder needs, but by resisting the normalization of partial or inaccessible explanations for those with the most at stake.

\subsection{Social Inequality}

Explanation practices are shaped by and can reinforce broader social hierarchies. Explanations often function as curated narratives that reflect institutional priorities more than user empowerment, where users are frequently given just enough information to comply with a decision, but not enough to challenge it \cite{eubanks2018automating}. Without critical reflection, explainability becomes a tool of legitimation, not justice. Designers must be attentive to how explanation choices reflect and reproduce systemic inequality in sociotechnical settings \cite{selbst2019fairness, barocas2020hidden, smart2024beyond}.

\subsection{Accountability and Governance}

Designing explanations is not just about making models interpretable but about making systems accountable \cite{doshi2017accountability, kim2020transparency, novelli2024accountability}. Explanations should help users understand model outputs \textit{and} surface the values, assumptions, and trade-offs embedded in the broader sociotechnical system \cite{ehsan2020human, alpsancar2024explanation}.

Accountability has become a matter of governance rather than simply an ethical ideal \cite{mokander2022algorithmic, lechterman2022concept, atoum2025revolutionizing}. As AI systems increasingly mediate access to housing, credit, healthcare, and public services, governments and institutions are beginning to impose regulatory frameworks to ensure these systems remain transparent, fair, and contestable. Recent policy developments such as the EU AI Act \cite{veale2021demystifying}, Canada's AIDA (Artificial Intelligence and Data Act) \cite{muhammad2023demystifying}, and algorithmic auditing mandates in cities like New York \cite{mokander2022us} reflect a growing consensus: explanations must not only serve users, but also serve governance functions by enabling oversight, documentation, and redress.

In this context, explanation is part of a broader infrastructure for institutional accountability beyond being a technical or cognitive aid. This includes designing explanations that can support audits \cite{zhang2022explainable}, creating documentation that traces design decisions, surfacing model uncertainties, and anticipating failure modes. Effective explanation design must therefore support accountability at multiple levels: interpersonal (e.g., enabling users to ask “why?”), organizational (e.g., allowing teams to trace and justify outcomes), and systemic (e.g., enabling regulators and public to interrogate high-stakes decisions). Explanation should not only clarify what a model did, but also make visible how responsibility is distributed, how decisions can be contested, and how harms can be redressed.

\section{Reframing XAI as Ethical Design Process}

Calls to reframe XAI as a human-centered process are increasingly common, but often underspecified in practice \cite{wolf2019explainability,wolf2020designing,ehsan2020human, ehsan2021expanding}. We have argued that explanation in AI systems must not be seen as a technical afterthought or post hoc rationalization, but as a consequential act of communication embedded within sociotechnical systems. Having made clear the stakes of ethical accounts in XAI, in this section, we will detail how our presented framework can account for ethical considerations.

The \textbf{WHO} dimension centers stakeholder diversity, not as a checklist of user roles (developer, validator, operator, subject), but as a reminder that explanations must navigate conflicting priorities and asymmetrical access to interpretive resources. Who gets a convincing explanation is a question of epistemic justice. A system that delivers rich, layered explanations to developers but vague justifications to subjects risks reinforcing existing inequities \cite{lazar2024legitimacy}.

The \textbf{WHAT} dimension highlights the contested nature of what is worth explaining. Whether one explains model inputs, outputs, processes, or counterfactuals is not a neutral design choice. These decisions shape how responsibility is distributed, which forms of error are visible, and whose knowledge is considered legitimate. Moreover, the drive for “faithfulness” often privileges model-internal logic over user-centered relevance, obscuring the social or policy context in which decisions unfold.

The \textbf{HOW} dimension, as discussed in the previous section, involves the translation of technical reasoning into formats legible to human interpreters. But rather than being the end of the design process, the delivery is part of an ongoing negotiation between intelligibility, persuasion, and control. Explanation design is not simply about usability; it is about shaping trust, setting boundaries of critique, and deciding who is entitled to question the system.

Most importantly, reframing XAI as an ethical design process is a call for epistemic humility, social equity, and accountability. It requires recognizing that explanation is always situated, produced under constraints, shaped by norms, and open to strategic misuse. This perspective challenges practitioners to ask not just whether explanations are accurate, but whether they are just, aid understanding, and whose understanding they prioritize \citep{lazar2024legitimacy}. And not just how they function, but how they might fail. In this light, XAI becomes less about ``making models explainable” and more about making such systems accountable through design choices that are explicit, reflective, and open to critique.

\section{Process Design in Action}

To concretize the idea of explanation as process designing, we walk through a structured example using the WHO, WHAT, and HOW framework. This thought experiment illustrates how explanation design unfolds under real-world constraints. Importantly, it also reveals how design decisions, though seemingly technical, are deeply ethical, reflecting choices about whose knowledge is privileged, what user assumptions are made, and which costs are deemed acceptable. Our example is meant as a guideline of implementing our framework and not a definitive reflection of exactly how the XAI process should look in deployment scenarios. 

\begin{tcolorbox}[colback=gray!5!white,colframe=gray!75!black,title=Case Study: Risk Prediction in Healthcare]
Consider a predictive model in a hospital context. The model estimates the 30-day readmission risk for patients with chronic heart failure. It is based on features like age, lab results, comorbidities, medications, and prior hospitalizations. 

The model already performs well on historical data. Now, it needs to support real-world decisions. How can an explanation model be constructed given the following points of interest?

\begin{itemize}
    \item The intended user is a clinician.
    \item The explanation must support actionable interventions.
    \item The hospital has limited resources to invest in explanation tooling.
\end{itemize}
\end{tcolorbox}

\subsection{WHO: Identifying the Stakeholder (and Who Gets Excluded)}

At first glance, the primary stakeholder is the clinician, an \textit{Operator} making treatment decisions based on model outputs. This classification guides the design process:

\begin{itemize}
    \item \textit{Context:} A clinician receives a high-risk alert for a patient and must respond quickly.
    \item \textit{Expertise:} Deep medical training, limited familiarity with ML.
    \item \textit{Goal:} Understand the prediction well enough to decide whether to intervene.
\end{itemize}

%\textit{Critical reflection:} While centering the clinician appears reasonable, this decision already narrows the ethical scope. Patients, those directly affected, are excluded from the explanation loop. Moreover, clinicians are not a homogenous group: a senior specialist and an overworked resident might have radically different capacities to engage with model explanations. By flattening ``the user” into a single archetype, design risks ignoring intra-stakeholder variance and reinforcing power asymmetries already present in clinical practice.

\textit{Ethical Reflection:} Centering the clinician in explanation design seems intuitive, given their decision-making role. However, this choice already shapes the epistemic boundaries of the system. By prioritizing the interpretive needs of the clinician, the design risks excluding the patient, the person most directly affected by the decision, from the explanation loop. This reflects a broader pattern of \textit{epistemic inequality}, where those with institutional authority are granted access to richer, more actionable explanations, while subjects of the model are offered only minimal or no insight into how decisions are made \cite{fricker2017evolving}. Furthermore, treating the clinician as a single, unified user archetype flattens important differences within the stakeholder group \cite{cabitza2017unintended}. A senior cardiologist and an overburdened medical resident may differ significantly in cognitive capacity, trust in automation, and time available to engage with an explanation. Without attending to these intra-stakeholder variances, design may unintentionally reinforce existing hierarchies in clinical workflows, privileging some users over others \cite{horsky2025cognitive}. This step thus demands a reflection on whose interpretive needs are being centered, whose are sidelined, and what institutional power dynamics are being reproduced in the process. % Ethically responsible explanation design begins by recognizing that stakeholder identification is not neutral is a normative decision that shapes access to understanding and participation, and ultimately, accountability.

\subsection{WHAT: Defining What Gets Explained (and What Gets Omitted)}

Designing the explanation content involves choosing from several dimensions:

\begin{itemize}
    \item \textit{Scope:} Local, patient-specific explanation is needed.
    \item \textit{Focus:} Feature attribution—what factors drove this specific prediction.
    \item \textit{Model access:} Full access permits use of model-specific techniques.
    \item \textit{Constraints:} Only low-compute methods are viable.
\end{itemize}

This points us to a feature attribution method like SHAP \cite{lundberg2017unified}, which can decompose the prediction into additive contributions of input features. It may seem like SHAP values are not amenable to understanding of our relevant stakeholder here (since physicians may not understand how SHAP works) but this should be resolved in the next ``HOW'' step where the focus is on how we communicate the explanation in an accessible format. 

\textit{Ethical Reflection:} While SHAP satisfies the design criteria, it also encodes assumptions that often go unquestioned: that causality can be approximated through additive feature contributions; that the model’s internal logic is a sufficient proxy for human understanding; and that technical constraints (e.g., compute) take precedence over epistemic ones (e.g., explanation completeness). These trade-offs may be acceptable, but only if they are surfaced, not silently embedded in the tooling. Explanation choices can conceal how social biases are encoded in model predictions. Methods like SHAP focus on which features contributed to a specific output, but they often overlook whether those features reflect underlying disparities in the training data. For example, if prior hospitalizations or medication access correlate with socioeconomic status, the explanation may quietly reproduce systemic inequities without surfacing them. Without careful dataset curation and proactive bias assessments, explanations risk legitimizing discriminatory patterns as neutral technical outputs \cite{antoniadi2021current, saraswat2022explainable, albahri2023systematic}. Therefore, explanation design content must go beyond technical fidelity to actively interrogate how social hierarchies are embedded in both data and model logic. What is left unexplained matters for accountability. When explanations focus solely on local feature effects, they may satisfy transparency checkboxes but fall short of supporting audits, redress, or governance. % Ethically robust explanation design must therefore go beyond technical interpretability to consider what is needed for institutional and regulatory accountability.

\subsection{HOW: Communicating the Explanation (and Whose Cognitive Fit We Privilege)}

Finally, we decide how to communicate these explanations effectively, tailoring the format to the stakeholder’s needs and context. In this case, we have our clinician who operates in a high-stakes, time-sensitive environment and needs to interpret the explanation quickly and confidently. 

\begin{itemize}
    \item Clinicians are accustomed to structured health data (e.g., lab reports, vitals) and often rely on both data visualizations
    \item The SHAP method we chose is amenable to creating visualizations. 
\end{itemize}

Given the clinician’s workflow and cognitive constraints, a ranked bar chart of SHAP values may be appropriate:

\begin{quote}
\textit{Top drivers: Recent ER admission (+0.24), Low sodium (+0.18), High blood pressure (+0.15)}
\end{quote}

To enhance usability, we might supplement this with a natural language summary:
\begin{quote}
\textit{“This patient is predicted to be high-risk primarily due to a recent ER visit, low sodium levels, and multiple prior readmissions.”}
\end{quote}

\textit{Ethical Reflection:} While such explanations may appear intuitive, they often present the model’s perspective as fact rather than construction. Moreover, designing for ``cognitive fit” can reinforce existing power imbalances, optimizing for what the high-status stakeholder finds legible, rather than empowering broader participation in contesting the model's decision. We must also contend with the risk of false confidence: clinicians may over-trust these explanations without being aware of their limitations, such as missing confounders, spurious correlations, or non-causal reasoning paths. Ethically responsible communication must therefore also include what the model cannot know, how it might fail, and where caution is warranted.

\subsection{Further Notes}  

From an ethical perspective on accountability and governance, explanation should not be treated as a one-time output but as part of an ongoing system of oversight and responsibility \cite{smith2020no}. Once an explanation is delivered, it must be possible to trace how it was generated, how it influenced decisions, and what outcomes it contributed to. This requires concrete institutional practices such as:

\begin{itemize}
    \item \textbf{Documentation:} To clearly state each explanation method's assumptions, limitations, and intended use cases \cite{bunn2020working}.
    \item \textbf{Evaluation:} The XAI methods being used should always be evaluated and such evaluation must be available on request to all stakeholders for greater transparency \cite{mohseni2021multidisciplinary, agarwal2022openxai, nauta2023anecdotal}.
    \item \textbf{Feedback Mechanisms:} To allow clinicians to flag confusing or misleading explanations. Such feedback is also often useful for improving the XAI system itself \cite{forster2025tell}.
    \item \textbf{Audit Trails:} To log when and how explanations are accessed and used in clinical decisions \cite{bagweexplainable}.
\end{itemize}

Without such support, explanations risk offering performative transparency, appearing interpretable while failing to enable contestation, accountability, or institutional learning. Ethical XAI design must therefore embed explanations into broader workflows of institutions.

%%%%%%%%%%%%%%%%%%%%%%%%%%%%%%%%%%%%%%%%%%%%%%%%%%%%%%%%%%%%%%%%%%%%%%%%%%%%%%%%%%%%%%%%%%%%%%%%%%%%%%%%%%%%%%%%%%%%%%%%%%%%%%%%%%%%%%%%%%%%

\section{Conclusion and Future Work}

This paper proposed a structured framework for explanation design in XAI, grounded in stakeholder-centered inquiry and ethical reflection. By centering the questions of \textit{who} an explanation is for, \textit{what} it should convey, and \textit{how} it should be delivered, we aim to reframe XAI as a design challenge embedded in social context, not just a technical problem of model introspection. Rather than prescribing fixed rules, this framework offers a foundation for creating explanation systems that are more adaptive and attentive to real-world use. However, there still remain many directions for future work. 

\textit{First}, empirical validation must go beyond traditional metrics, e.g., satisfaction, trust scores \cite{alam2021examining,brdnik2023assessing} and engage with the lived experiences of stakeholders using XAI in practice \cite{hoffman2023measures}. As recent work in HCI shows, explanation systems often produce unintended effects, such as misplaced trust, cognitive overload, or institutional drift from original design intent \cite{bhatt2020explainable, liao2022connecting}. Ethnographic methods \cite{wolf2019explainability} and participatory evaluation \cite{hashmati2024explainable} will be essential to assess not only usability, but also which groups benefit from an explanation strategy, and which may be excluded, confused, or burdened by it.

\textit{Second}, there is a pressing need for design tools and workflows that are both actionable and ethically robust. Toolkits for explanation should support inclusive participation during prototyping, as well as throughout deployment, and iteration \cite{ehsan2021expanding}. Such methods would ideally include checklists that flag risks of epistemic injustice \cite{fricker2017evolving}, and interfaces that let users question or reject explanations \cite{sovrano2021philosophy}. However, simplification can also conceal harm: participatory design must avoid becoming a token gesture, and future work should develop critical reflection practices that help teams recognize when efficiency or resource constraints are quietly narrowing whose needs are being served.

\textit{Third}, explanation design must engage more directly with the material constraints and political realities that shape real-world AI systems. Resource limitations are not just technical boundaries; they reflect institutional priorities and power dynamics \cite{edwards2017slave, edwards2018enslaving}. As emerging regulation makes clear, technology will get increasingly tied to compliance and auditability \cite{mokander2022algorithmic}. Future research should explore how explanation quality is distributed across institutions (e.g., well-resourced vs underfunded hospitals) and stakeholders (e.g., clinicians vs patients), and investigate not only how to adapt explanations under constraint, but why those constraints exist, who imposes them, and whether they can or should be contested \cite{birhane2022power, birhane2022forgotten, farrow2023possibilities}.

Ultimately, this framework invites the reader to rethink what explanation is and what it is for. Explanation shouldn’t be treated as a static output or interface widget, but as an evolving negotiation among values, stakeholders, and institutions. Explanation should be understood not as a final product, but as part of a broader system of interaction, intervention, and responsibility. Future work should treat explanation as infrastructure that supports understanding, action, and accountability across real-world settings.

\section{Acknowledgments}

The authors would like to express their sincere gratitude to the anonymous reviewers for their valuable comments and constructive feedback. We also acknowledge the support of Pioneer Centre for AI under Grant No. P1, whose funding made this research possible.

\bibliography{aaai25}

\end{document}